\documentclass[12pt,preprint]{aastex}

\def\be{\begin{equation}}
\def\ee{\end{equation}}
\def\ba{\begin{eqnarray}}
\def\ea{\end{eqnarray}}
\newcommand{\degree}{\ensuremath{^\circ}}

\newcommand{\msun}{\ifmmode\mbox{M}_{\odot}\else$\mbox{M}_{\odot}$\fi}
\newcommand{\rsun}{\ifmmode\mbox{R}_{\odot}\else$\mbox{M}_{\odot}$\fi}
\newcommand{\degrees}{\ifmmode^{\circ}\else$^{\circ}$\fi}
\newcommand{\amin}{\ifmmode^{\prime}\else$^{\prime}$\fi}
\newcommand{\asec}{\ifmmode^{\prime\prime}\else$^{\prime\prime}$\fi}

\shorttitle{Two Millisecond Pulsars}
\shortauthors{Deneva et al.}

\begin{document}

\title{Two Millisecond Pulsars Discovered by the PALFA Survey and a Shapiro Delay Measurement}

\author{
J. S. Deneva\altaffilmark{1},
P. C. C. Freire\altaffilmark{2},
J. M. Cordes\altaffilmark{3},
A. G. Lyne\altaffilmark{4},
S. M. Ransom\altaffilmark{5},
I. Cognard\altaffilmark{6},
F. Camilo\altaffilmark{7},
D. J. Nice\altaffilmark{8},
I. H. Stairs\altaffilmark{9},
B. Allen\altaffilmark{10},
N. D. R. Bhat\altaffilmark{11},
S. Bogdanov\altaffilmark{7},
A. Brazier\altaffilmark{3},
D. J. Champion\altaffilmark{2},
S. Chatterjee\altaffilmark{3},
F. Crawford\altaffilmark{13},
G. Desvignes\altaffilmark{2},
J. W. T. Hessels\altaffilmark{14},
F. A. Jenet\altaffilmark{15},
V. M. Kaspi\altaffilmark{12},
B. Knispel\altaffilmark{10},
M. Kramer\altaffilmark{2},
P. Lazarus\altaffilmark{2,12},
J. van Leeuwen\altaffilmark{15},
D. R. Lorimer\altaffilmark{16},
R. S. Lynch\altaffilmark{12},
M. A. McLaughlin\altaffilmark{16},
P. Scholz\altaffilmark{12},
X. Siemens\altaffilmark{17},
B. W. Stappers\altaffilmark{4},
K. Stovall\altaffilmark{15},
A. Venkataraman\altaffilmark{1}
}

\altaffiltext{1}{Arecibo Observatory, HC3 Box 53995, Arecibo, PR
00612, USA}

\altaffiltext{2}{Max-Planck-Institut f\"{u}r Radioastronomie, D-53121
Bonn, Germany}

\altaffiltext{3}{Astronomy Department, Cornell University, Ithaca, NY
14853, USA}

\altaffiltext{4}{Jodrell Bank Centre for Astrophysics, University of 
Manchester, Manchester M13 9PL, UK}

\altaffiltext{5}{National Radio Astronomy Observatory,
Charlottesville, VA 22903, USA}

\altaffiltext{6}{Laboratoire de Physique et Chimie de l'Environnement et
de l'Espace, LPC2E, CNRS et Universit\'{e} d'Orl\'{e}ans, and Station de
radioastronomie de Nan\c{c}ay, Observatoire de Paris, France}

\altaffiltext{7}{Columbia Astrophysics Laboratory, Columbia
University, New York, NY 10027, USA}

\altaffiltext{8}{Department of Physics, Lafayette College, Easton, PA
18042, USA}

\altaffiltext{9}{Department of Physics and Astronomy, University of
British Columbia, 6224 Agricultural Road, Vancouver, BC V6T 1Z1, Canada}

\altaffiltext{10}{Albert-Einstein-Institut, Max-Planck-Institut f\"{u}r
Gravitationsphysik, D-30167 Hannover, Germany, and Institut f\"{u}r
Gravitationsphysik, Leibniz Universit\"{a}t Hannover, D-30167 Hannover,
Germany}

\altaffiltext{11}{Swinburne University, Center for Astrophysics and
Supercomputing, Hawthorn, Victoria 3122, Australia}

\altaffiltext{12}{Department of Physics, McGill University, 3600 rue Universite, Montreal, QC H3A 2T8, Canada}

\altaffiltext{13}{Department of Physics and Astronomy, Franklin and
Marshall College, P.O. Box 3003, Lancaster, PA 17604, USA}

\altaffiltext{14}{ASTRON, the Netherlands Institute for Radio
Astronomy, Postbus 2, 7990 AA, Dwingeloo, The Netherlands and
Astronomical Institute ``Anton Pannekoek,'' University of Amsterdam,
Science Park 904, 1098 XH Amsterdam, The Netherlands}

\altaffiltext{15}{Center for Gravitational Wave Astronomy, University
of Texas at Brownsville, Brownsville, TX 78520, USA}

\altaffiltext{16}{Department of Physics, West Virginia University,
Morgantown, WV 26506, USA}

\altaffiltext{17}{Physics Department, University of Wisconsin at
Milwaukee, Milwaukee, WI 53211, USA}

\begin{abstract}
We present two millisecond pulsar discoveries from the PALFA survey of the Galactic plane with the Arecibo telescope. PSR~J1955+2527 is an isolated pulsar with a period of 4.87~ms, and PSR~J1949+3106 has a period of 13.14~ms and is in a 1.9-day binary system with a massive companion.
Their timing solutions, based on 4 years of timing measurements with the Arecibo, Green Bank, Nan\c{c}ay and Jodrell Bank telescopes, allow precise determination of spin and astrometric parameters, including precise determinations of their proper motions. For PSR~J1949+3106, we can clearly detect the Shapiro delay. From this we measure the pulsar mass to be $1.47^{+0.43}_{-0.31}$~\msun, the companion mass to be $0.85^{+0.14}_{-0.11}$~\msun\ and the orbital inclination to be $i = 79.9_{+1.6}^{-1.9}$ degrees, where uncertainties correspond to $\pm 1$-$\sigma$ confidence levels. With continued timing, we expect to also be able to detect the advance of periastron for the J1949+3106 system. This effect, combined with the Shapiro delay, will eventually provide very precise mass measurements for this system and a test of general relativity.

\end{abstract} 

\section{Introduction}

In this paper, we discuss timing results for two millisecond pulsars (MSPs) discovered by the PALFA Consortium\footnote{\tt http://www.naic.edu/alfa/pulsar} with the Arecibo telescope. The PALFA survey of the Galactic plane and follow-up observations of new discoveries are motivated by the wide applications of pulsar timing in exploring the composition, internal structure, and magnetospheric state of neutron stars. Millisecond pulsars in particular tend to be extremely stable rotators, which can be used to address a variety of problems in fundamental physics and astrophysics. One such outstanding problem is detecting gravitational waves and studying the properties of gravitational wave radiation from various types of sources and various epochs in the Universe's lifetime. Timing observations of MSPs in binary systems can be used to estimate the pulsar and companion masses via measuring post-Keplerian binary parameters. Neutron star mass measurements allow us to constrain the equation of state (EoS) of matter at densities larger than that of an atomic nucleus.

Until recently, all precise measurements of neutron star masses fell within a narrow range around 1.4~\msun, the Chandrasekhar limit. However, recent, precise measurements of the masses of some MSPs showed that they can have significantly higher masses. PSR~J1903+0327 \citep{Champion08}, the first MSP found in the PALFA survey, has a mass of $1.67 \pm 0.02~\msun$ (99.7 \% confidence limit, \citealt{Freire11}) which is significantly above the Chandrasekhar limit. PSR J1614$-$2230 was found to have a mass of $1.97 \pm 0.04~\msun$ \citep{Demorest10}. 
These high masses rule out many EoSs for matter at densities higher than that of the atomic nucleus. In particular, the mass measurement for J1614$-$2230 rules out or highly constrains most proposed hyperon or boson EoSs (\citealt{Glendenning98}, \citealt{Lackey06}, \citealt{Schulze06}, \citealt{Lattimer07}).

In this paper, we describe two MSPs discovered by the PALFA survey, which uses the Arecibo telescope and the seven-beam ALFA receiver \citep{Cordes06}. In addition to presenting full timing solutions for the two pulsars, we explore what intrinsic or extrinsic effects account for the overall TOA uncertainty in both cases. One of the two discoveries reported here is in a nearly edge-on binary system with a massive companion, allowing measurement of the Shapiro delay and consequently estimation of the pulsar and companion masses.

\section{Observations and Data Reduction}

PSR J1949+3106 and PSR J1955+2527 were discovered in PALFA survey data taken in 2006 and processed by the Cornell search pipeline \citep{Deneva09}. The PALFA survey is ongoing since 2004 and uses the seven-beam Arecibo L-band Feed Array (ALFA) to search for pulsars in low Galactic latitudes ($|b| < 5\degree$) in the portions of the Galactic plane visible from Arecibo, $32\degree < l < 77\degree$ and $168\degree < l < 214\degree$. PALFA has discovered 97 new pulsars, including 15 MSPs, as of July 2012 (\citealt{Cordes06}, \citealt{Lorimer06}, \citealt{Champion08}, \citealt{Deneva09}, \citealt{Knispel10}, \citealt{Knispel11}, \citealt{Nice12}, \citealt{Crawford12}). There are currently 23 pulsars in the ATNF catalog\footnote{\texttt{http://www.atnf.csiro.au/research/pulsar/psrcat}} \citep{ATNF} with $P < 15$~ms and $|b| < 5$\degree\ that are not in globular clusters. Of those, eight have dispersion measures (DMs) $ > 100$~pc~cm$^{-3}$, and only one has $\rm DM > 200$~pc~cm$^{-3}$. That pulsar is J1903+0327 \citep{Champion08}, an eccentric binary MSP discovered by the PALFA survey. The two discoveries we report, J1949+3106 and J1955+2527, have $\rm DM = 164$~pc~cm$^{-3}$ and 209~pc~cm$^{-3}$, respectively. This makes the latter only the second Galactic plane MSP known with $\rm DM > 200$~pc~cm$^{-3}$. \cite{Crawford12} present four new MSPs with $DM > 100$~pc~cm$^{-3}$, also found by PALFA. One of these four pulsars has $DM = 249$~pc~cm$^{-3}$. Along with the high DMs of J1903+0327 and J1955+2527, this shows that the PALFA survey is probing deeper and finding MSPs whose detectability with earlier instruments may have been hampered by dispersion smearing.

In 2006, the PALFA survey used the Wideband Arecibo Pulsar Processor (WAPP) backends \citep{Dowd00} with 100~MHz bandwidth per beam, two summed polarization channels, 256 lags, and 64~$\mu$s sampling time \citep{Cordes06}. Since 2009 the survey uses the newer PDEV/Mock spectrometer\footnote{\texttt{http://www.naic.edu/∼phil/hardware/pdev/usersGuide.pdf}}, which allows data to be taken across the entire 300~MHz bandwidth of the ALFA receiver for each beam, thus increasing sensitivity by a nominal factor of $\sqrt{3}$. This observation setup uses two summed polarizations, 960 channels, and a samping time of 65.5~$\mu$s.

PSR J1949+3106 and PSR J1955+2527 were found in a conventional Fourier-based periodicity search of WAPP data and were not detected in a single pulse search. Timing observations of the pulsars began initially with the Green Bank Telescope. After a dedicated timing program was established, higher-sensitivity regular timing observations were done with the Arecibo, Nan\c cay, and Jodrell Bank telescopes.

\subsection{Timing Observations}

Green Bank timing observations used the S-band receiver and the Pulsar Spigot backend \citep{Kaplan05} with a total bandwidth of 800~MHz centered on 1890~MHz, 1024 lags, and 81.92~$\mu$s sampling time. Because of strong and persistent RFI, the lowest 200~MHz of the band is excluded, leaving 600~MHz of effective bandwidth centered on 1950~MHz. The Arecibo observations used the L-band Wide receiver and WAPP correlator backends. The four WAPPs covered adjacent 50~MHz bands with 512 lags each, centered at 1420, 1470, 1520, and 1570~MHz. The sampling time for Arecibo observations was 64~$\mu$s. GBT and Arecibo observations were made in search mode, with two summed polarizations and continuous data streams recorded for off-line processing. After data collection, the outputs of the spectral channels were folded at the predicted topocentric period of the pulsar, shifted to compensate for dispersive delays, and summed to produce a total intensity pulse profile. 

Observations with the Lovell telescope covered a 300-MHz band centered on 1500 MHz. A digital filterbank with a channel width of 0.5~MHz was used in a synchronous integration mode, in which the data acquisition is synchronous with the pulsar period. Baseband data were collected for the duration of a time bin (1/1024 of a pulse period). The power spectrum of these data is then calculated and added to the cumulative spectrum for that bin. After shifting to compensate for dispersive delays, the data were summed to produce a total intensity profile. This setup results in a sampling time of 12.83~$\mu$s for J1949+3106 and 4.76~$\mu$s for J1955+2527. 

Nan\c{c}ay observations were performed with the Berkeley-Orl\'eans-Nan\c{c}ay instrument \citep{Cognard09} and used coherent dedispersion performed by Graphics Processing Units (GPUs) in 4 MHz channels over a total bandwidth of 128~MHz centered at 1400~GHz. For each pulsar, all the observations at each frequency were integrated to produce a template used to derive topocentric times of arrival (TOAs) following a standard $\chi^2$ fit in the frequency domain \citep{Taylor92}.

We used the TEMPO2 software package \citep{Tempo06} to perform least-squares fitting of various pulsar parameters by minimizing the square of the differences between expected and measured TOAs. We directed TEMPO2 to fit for constant offsets between TOA sets from different observatories by bracketing each TOA set with JUMP statements. We scaled the TOA uncertainties (via EFAC statements, Table~\ref{tab_toaerror}) so that for each TOA set the ratio of $\chi^2$ to the number of degrees of freedom is close to unity. TEMPO2 applies clock corrections based on the location of each observatory and Earth rotation data to convert TOAs to terrestrial time (TT\footnote{From Bureau International des Poids et Mesures (BIPM)}). Conversion from TT to coordinated barycentric time (TCB) was done using the DE405 Solar System ephemeris \citep{Standish98}. For J1949+3106, we used the Damour \& Deruelle (henceforth DD) orbital model \citep{DD86}, with the Shapiro delay parameterized as in Freire \& Wex (2010). The resulting best-fit parameters and other derived pulsar parameters are listed in Table \ref{tab_pulsars}. There are no significant trends in the timing residuals; this implies that at the current timing precision, the ephemerides presented in Table \ref{tab_pulsars} describe the TOAs. Both pulsars have now been timed for four years. This has allowed very precise measurements of the spin and astrometric parameters, in particular precise determinations of the proper motions of these two new objects, to be discussed below.

\subsection{Polarimetry}

\begin{figure}
\begin{center}
\includegraphics[scale=0.75, angle=-90]{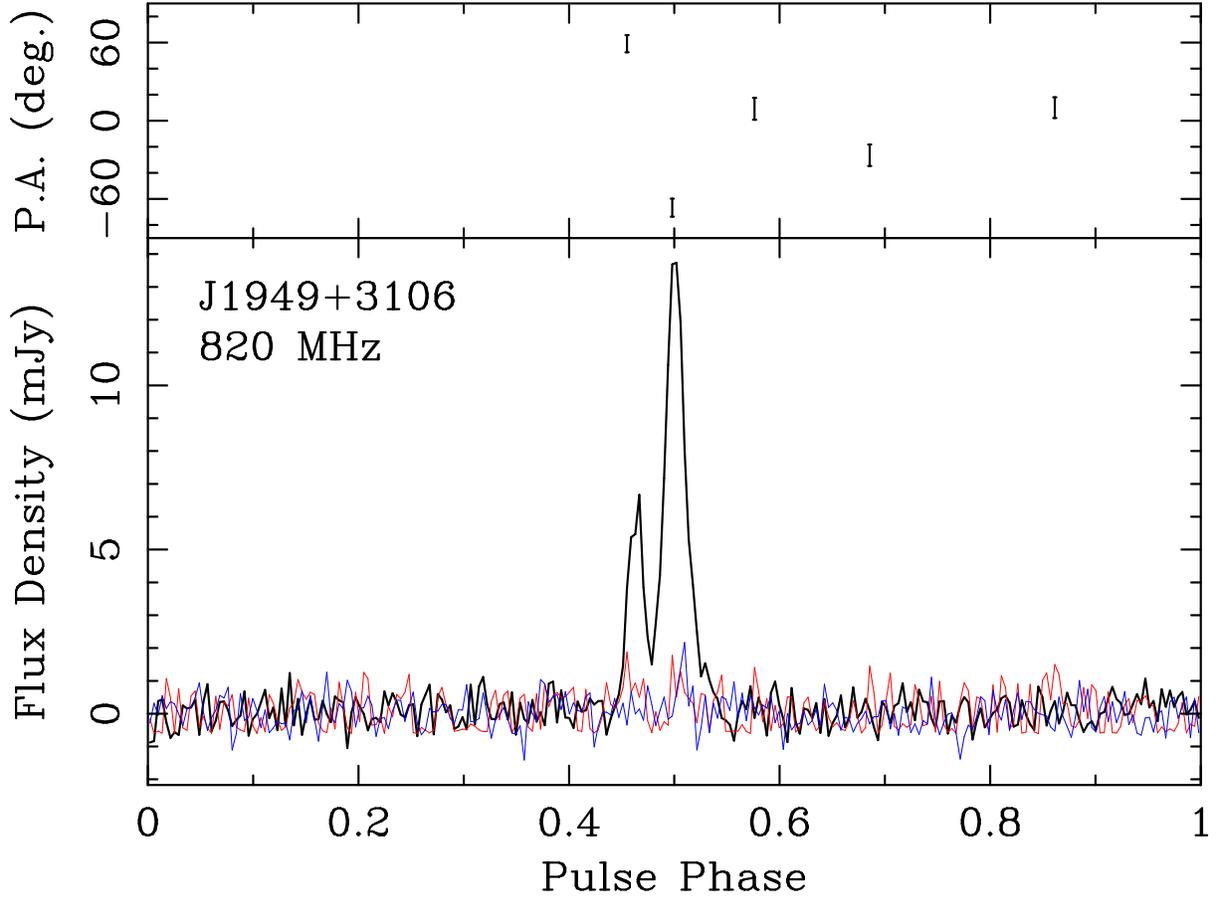}
\caption[]{\small Polarimetric pulse profiles for PSR~J1949+3106 at 820\,MHz.  This is
based on a 25 minute observation obtained at the GBT with GUPPI \citep{Demorest10},
and is displayed with 256 bins.  In the bottom plot, the black trace
corresponds to total intensity, while the red and blue lines correspond
to linear and circular polarization, respectively.  In the top plot,
the position angle of linear polarization (PA) is plotted for bins in the the linear polarization profile with signal-to-noise ratio $>3$. Bins outside the main pulse profile that exceed this threshold are likely due to statistical fluctuations or imperfectly cleaned RFI.}\label{fig_1949polGBT}
\end{center}
\end{figure}

\begin{figure}
\begin{center}
\includegraphics[scale=0.75, angle=-90]{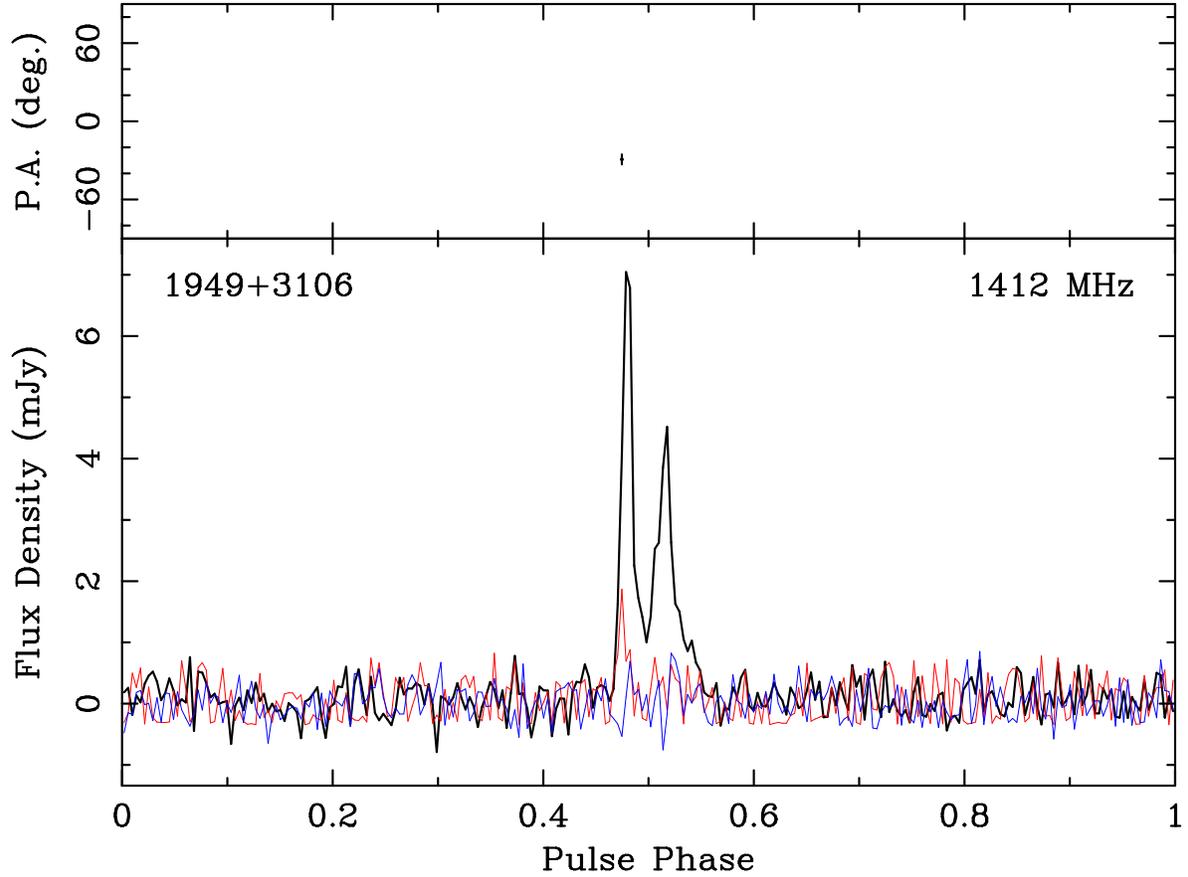}
\caption[]{\small Polarimetric pulse profiles for PSR~J1949+3106 at 1412\,MHz.  This is
based on a 15 minute observation obtained at the Arecibo telescope with the ASP \citep{Demorest07}, and is displayed with 256 bins.  In the bottom plot, the black trace
corresponds to total intensity, while the red and blue lines correspond
to linear and circular polarization, respectively.  In the top plot,
the position angle of linear polarization (PA) is plotted for bins in the the linear polarization profile with signal-to-noise ratio $>3$.}\label{fig_1949polAO}
\end{center}
\end{figure}

In order to measure the polarization characteristics of PSR~J1949+3106, we observed it in coherent dedispersion mode at two different frequencies. Our 25 minute GBT observation used GUPPI\footnote{https://wikio.nrao.edu/bin/view/CICADA/GUPPiUsersGuide} to sample a bandwidth of 200~MHz centered on 820~MHz. We also observed the pulsar for 15 minutes at Arecibo using the ASP backend \citep{Demorest07} with a bandwidth of 24~MHz centered on 1412~MHz. The data were analyzed in standard fashion with PSRCHIVE (Hotan et al.\ 2004), and the resulting calibrated full-Stokes pulse profiles are shown in Figure~\ref{fig_1949polGBT} and Figure~\ref{fig_1949polAO}. 

\begin{figure}
\begin{center}
\includegraphics[scale=0.8,angle=0]{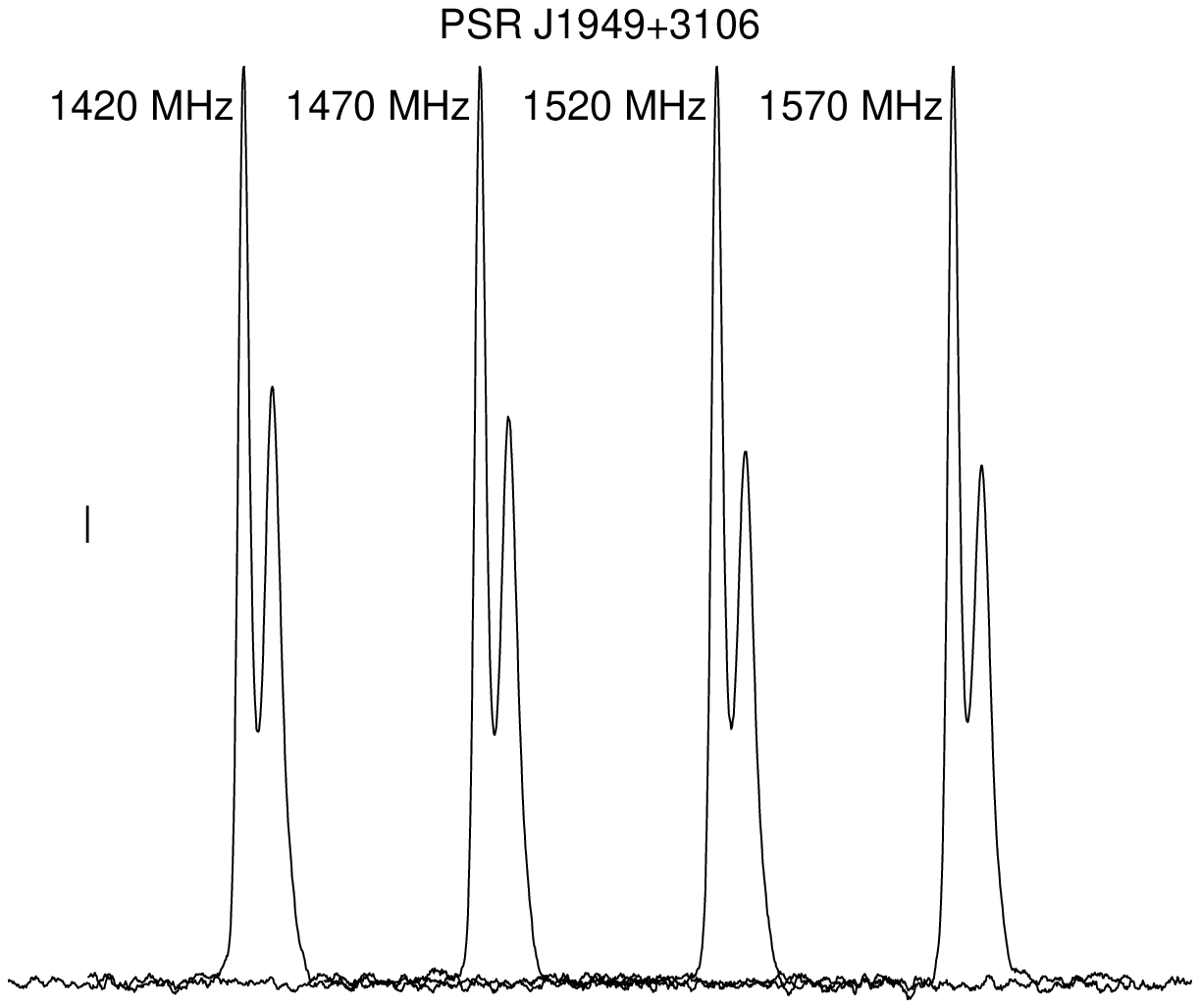}
\includegraphics[scale=0.7,angle=0]{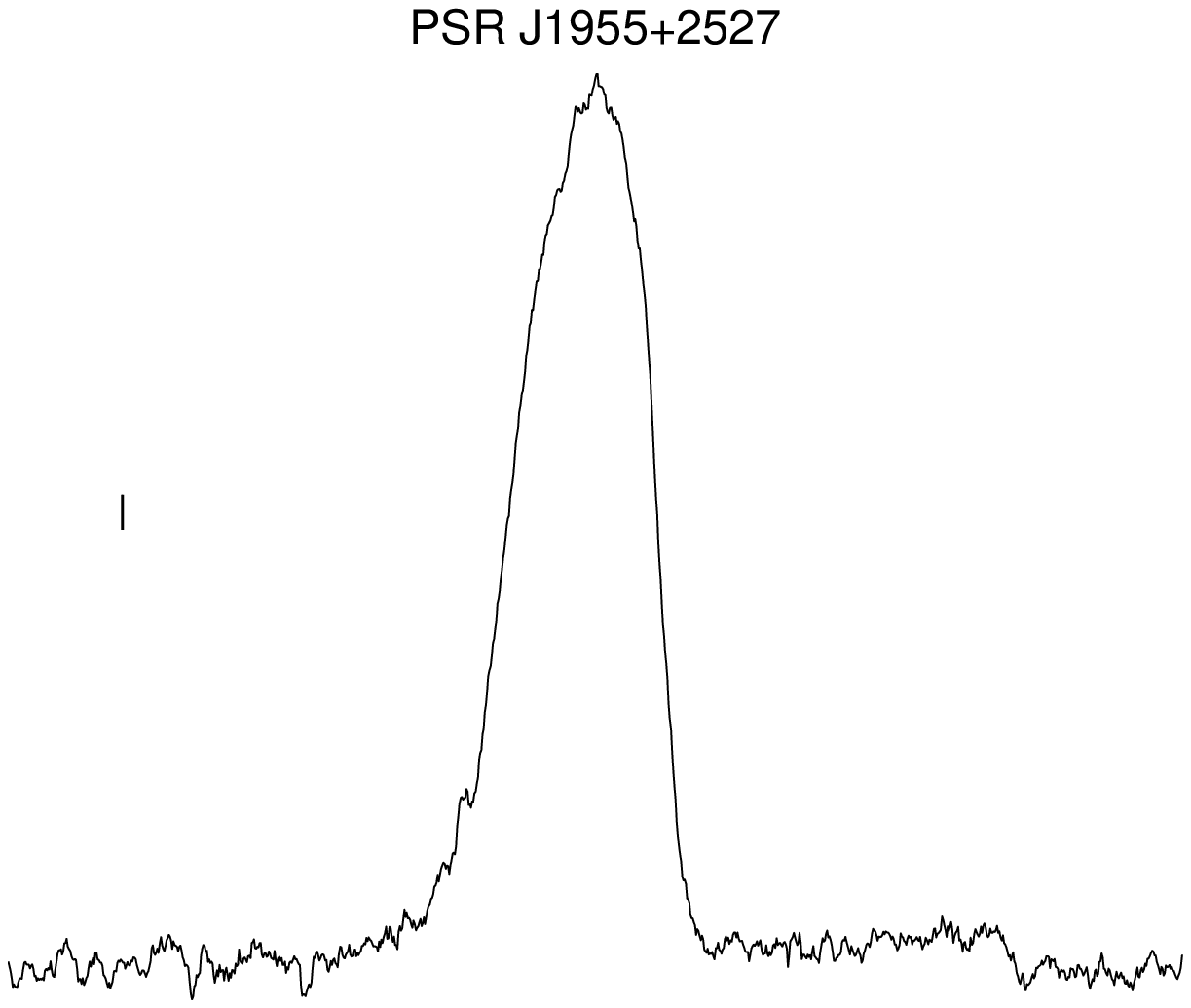}
\caption[]{\small Top: Pulse profiles for J1949+3106 from the four 50~MHz WAPP bands used in Arecibo timing observations. The two peaks have opposite relative strengths at L-band compared to 820~MHz (Fig.~\ref{fig_1949polGBT}). Profile evolution is evident across the four WAPP bands: the trailing peak decreases in strength relative to the leading peak with increasing frequency. Bottom: Pulse profile for J1955+2527 from the 50~MHz WAPP band centered on 1420~MHz. Folded profiles from all Arecibo observations were aligned and averaged to produce these plots. For both pulsars, profiles use 1024 bins and have been normalized to the same height on the arbitrary vertical axis. The thickness of the line on the left middle of each plot corresponds to the width of a time bin.}\label{fig_profs}
\end{center}
\end{figure}

At 820~MHz, there is little to no circular polarization in the second component of the pulse profile, or linear polarization in both components. This is somewhat unusual since pulsars are among the most polarized radio sources and individual pulsars can be up to 100\% polarized (e.g. \citealt{Han09}). At 1412~MHz, there is some linear polarization in the first component, and the second seems unpolarized. The rotation measure is unconstrained in either observation.  Comparing the two profiles, it is clear that the second component has a steeper spectrum than the first. This is confirmed by the observed profile evolution with frequency across the four WAPP bands used in Arecibo timing observations (Figure~\ref{fig_profs}). Assuming a 20\% uncertainty, the measured period-averaged flux density is $0.39\pm0.08$\,mJy for the 820~MHz GBT observation and $0.23\pm0.05$~mJy for the 1412~MHz Arecibo obsservation, giving a spectral index of $\alpha = -0.97$.

\begin{figure}
\begin{center}
\includegraphics[scale=0.75, angle=-90]{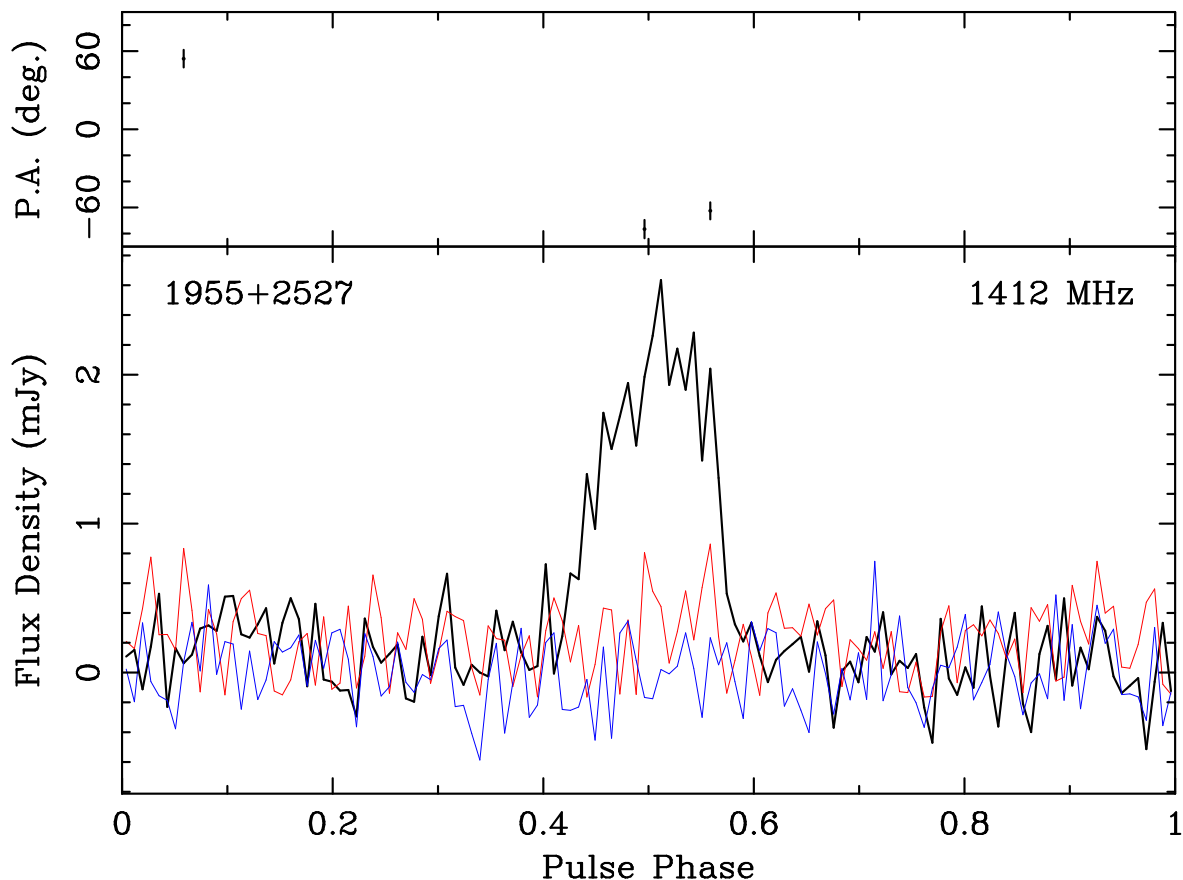}
\caption[]{Polarimetric pulse profiles for PSR~J1955+2527 at 1412\,MHz.  This is
based on a 10 minute observation obtained at the Arecibo telescope with the ASP \citep{Demorest07}, and is displayed with 128 bins.  In the bottom plot, the black trace
corresponds to total intensity, while the red and blue lines correspond
to linear and circular polarization, respectively.  In the top plot,
the position angle of linear polarization (PA) is plotted for bins
in the the linear polarization profile with signal-to-noise ratio $>3$.
Bins outside the main pulse profile that exceed this threshold are likely due to statistical fluctuations or imperfectly cleaned RFI.}\label{fig_1955polAO}
\end{center}
\end{figure}

We observed PSR~J1955+2527 for 10 minutes using the same Arecibo ASP setup and data reduction method described above. The calibrated full-Stokes pulse profiles are shown in Figure~\ref{fig_1955polAO}. There is little to no linear polarization and no circular polarization in the pulse profile, and the rotation measure is not constrained. The period-averaged flux density at 1412~MHz is $0.28\pm0.06$~mJy.

\section{PSR J1955+2527}\label{sec_1955}

\begin{figure}
\begin{center}
\includegraphics[scale=0.6, angle=-90]{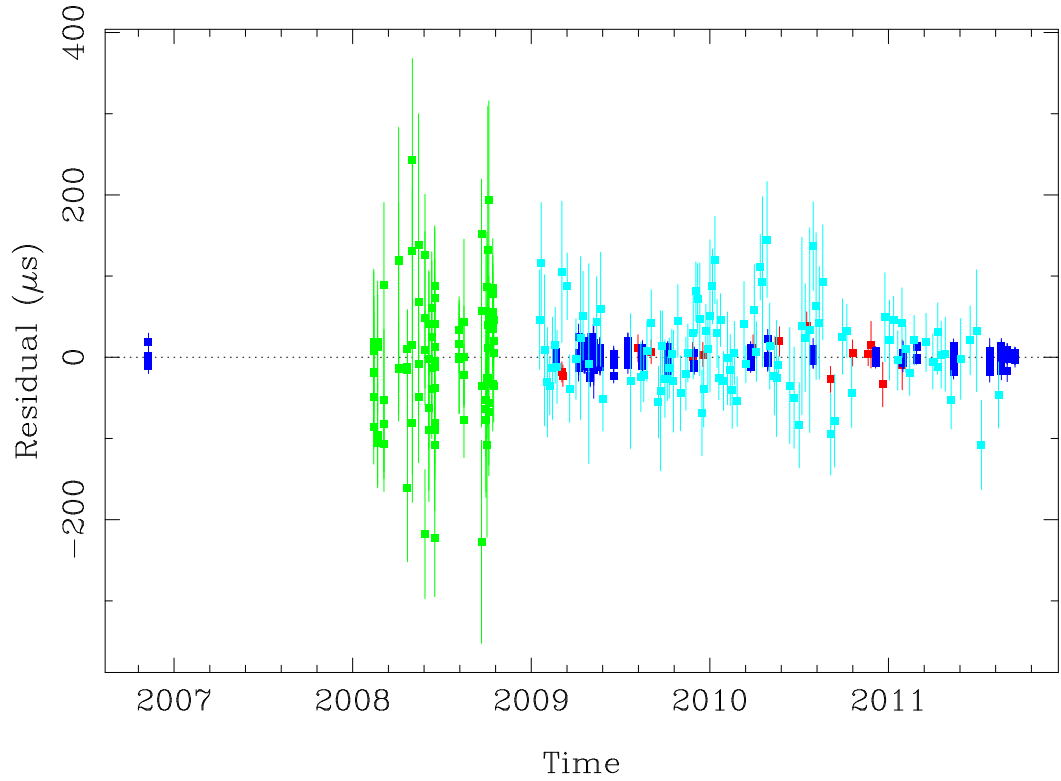}
\caption[]{Timing residuals vs. epoch for PSR J1955+2527. Green TOAs are from GBT observations, blue TOAs are from Arecibo observations, cyan TOAs are from Jodrell observations, and red TOAs are from Nan\c cay coherently dedispersed data.}\label{fig_1955resid}
\end{center}
\end{figure}

PSR~J1955+2527 is an isolated MSP with a period of 4.87~ms and dispersion measure of 209.97~pc~cm$^{-3}$. Figure~\ref{fig_profs} shows a folded pulse profile from an Arecibo timing observation. Table~\ref{tab_pulsars} summarizes the timing solution parameters, and Figure~\ref{fig_1955resid} shows timing residuals vs. epoch for the TOAs after removing the best-fit timing solution. 

With its overall post-fit RMS timing residual of 11.4~$\mu$s, J1955+2527 does not have high enough timing precision to be included in the pulsar sample used by Pulsar Timing Array projects to attempt detection of gravitational waves\footnote{\tt http://www.ipta4gw.org}. An RMS residual on the order of 100~ns or lower is required for that purpose. In Section~\ref{sec_errorbudget} we investigate the contributions of various intrinsic and extrinsic effects to the timing residuals of both J1955+2527 and J1949+3106. 

\section{PSR J1949+3106}

\begin{figure}
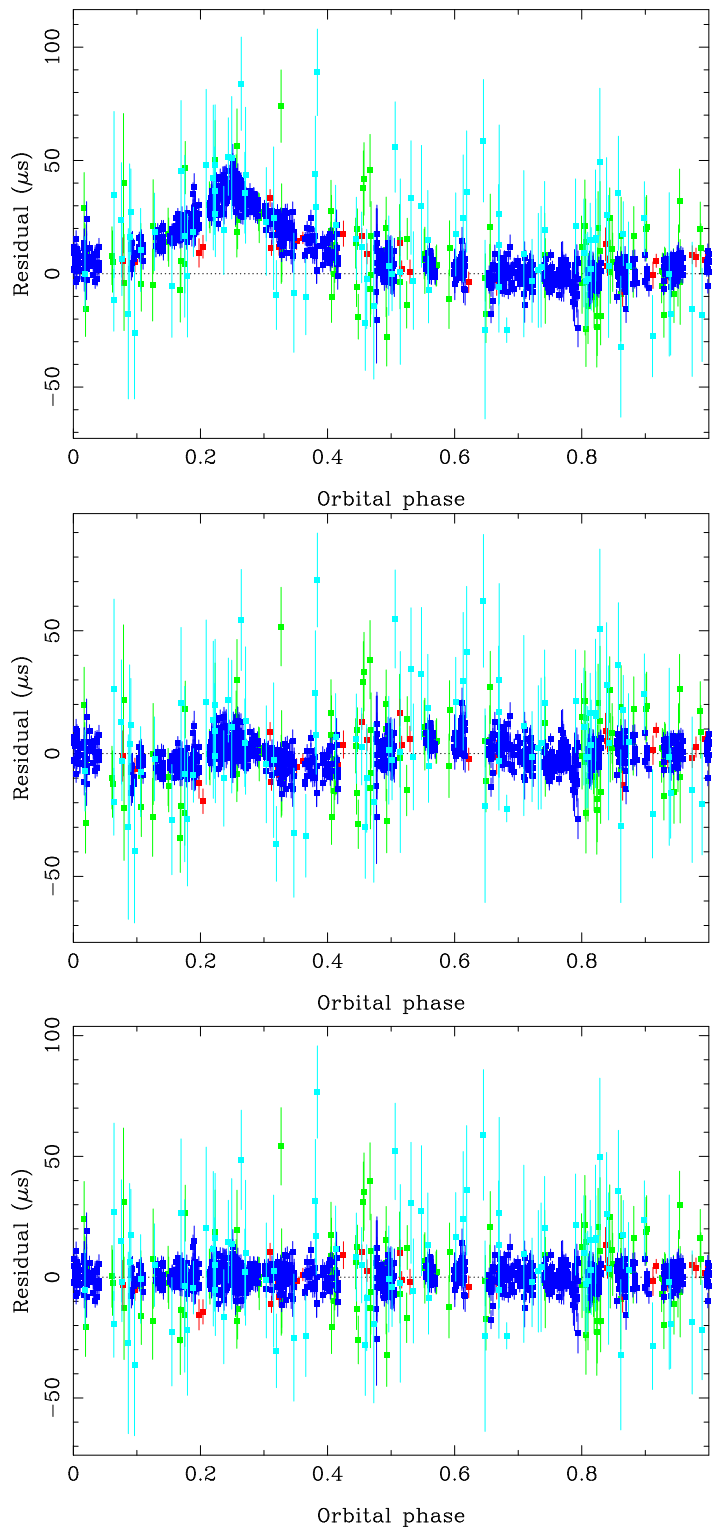

\begin{center}
\includegraphics[scale=0.4, angle=-90]{fig6a.ps}
\includegraphics[scale=0.4, angle=-90]{fig6b.ps}
\includegraphics[scale=0.4, angle=-90]{fig6c.ps}
\caption[]{Timing residuals vs. orbital phase for PSR J1949+3106 before (top) and after (bottom) fitting for Shapiro delay parameters. The middle plot shows the part of the Shapiro delay that is not absorbed by the fits of Keplerian parameters of the system. Green TOAs are from GBT observations, blue TOAs are from Arecibo observations, cyan TOAs are from Jodrell Bank observations, and red TOAs are from Nan\c cay coherently dedispersed data.}\label{fig_1949shapiro}
\end{center}
\end{figure}

PSR~J1949+3106 is an MSP with a period of 13.14~ms and DM of 164.13~pc~cm$^{-3}$. It is in a 1.95-day binary system, and because the orbital plane of the binary is inclined by $\sim 80\degrees$ as viewed from Earth, we are able to measure the Shapiro delay in the system and derive estimates of the pulsar and companion masses, and the orbital inclination angle. 
Table~\ref{tab_pulsars} summarizes the timing solution parameters for J1949+3106, and Figure~\ref{fig_1949shapiro} shows timing residuals vs.\ orbital phase for the TOAs used to obtain the solution. Figures~\ref{fig_1949polGBT}, \ref{fig_1949polAO}, and \ref{fig_profs} show folded pulse profiles for J1949+3106 at 820~MHz, 1412~MHz, and the four WAPP subbands, respectively.

\subsection{Keplerian parameters}
\label{sec:Keplerian}

The timing solution of any pulsar binary includes the orbital period $P_{\rm b}$, the projected semi-major axis of the orbit $x$ (typically expressed in light seconds), and parameters depending on the orbital eccentricity $e$, the time of passage through periastron $T_0$, and the longitude of the ascending node $\omega$. For PSR~J1949+3106 they are presented in Table~\ref{tab_pulsars}. From these, we can estimate the mass function of the system:
\be
f\left(m_{\rm p}, m_{\rm c} \right) = \frac{\left( m_{\rm c}~{\rm sin}~i\right)^3}{\left(m_{\rm p}+m_{\rm c} \right)^2} = \frac{4 \pi^2 x^3 c^3}{G P_{\rm b}^2},
\ee
where $m_{\rm p}$ is the mass of the pulsar, $m_{\rm c}$ is the companion mass, $i$ is the orbital inclination angle with respect to the plane of the sky, $c$ is the speed of light, and $G$ is the gravitational constant. For PSR~J1949+3106, we obtain $f = 0.10938(5)$\,\msun. Assuming that the pulsar has a mass close to the Chandrasekhar limit (1.4 \msun) and an orbital inclination $i = 90^\circ$, we obtain a minimum companion mass of 0.8 \msun. To disentangle the masses of the two components from the sine of the orbital inclination angle, we must be able to measure at least two post-Keplerian parameters (\S~\ref{sec:Shapiro}).


\subsection{Shapiro delay}
\label{sec:Shapiro}

One important goal of pulsar surveys in general is to discover pulsars whose properties can constrain the equation of state of neutron-star matter. Current proposed equations of state differ in their allowed ranges for pulsar rotation periods and masses. Therefore, two ways of constraining them are to find very fast-spinning ($P \lesssim 1$~ms) or very massive neutron stars ($m_{\rm p} \gtrsim 1.8 - 2.0$~\msun). While pulsar rotation periods are obtained immediately upon discovery from Fourier-based search algorithms, measuring pulsar masses is only possible for binary systems in which one or more post-Keplerian effects can be measured.

When the companion is between the pulsar and the Earth, pulses traveling from the pulsar to the Earth pass through the gravitational well of the companion and a relativistic delay is introduced in their arrival time as seen by the observer--the Shapiro delay (e.g. \citealt{Shapiro64}). Most orbital models parametrize the Shapiro delay in terms of the ``range'' $r = G c^{-3} m_{\rm c}$ and ``shape'' $s = {\rm sin}\ i$. Once we have fits for $r$ and $s$, we can solve for $m_{\rm p}$ based on the mass function of the system.



\begin{figure}
\begin{center}
\includegraphics[scale=0.7, angle=0]{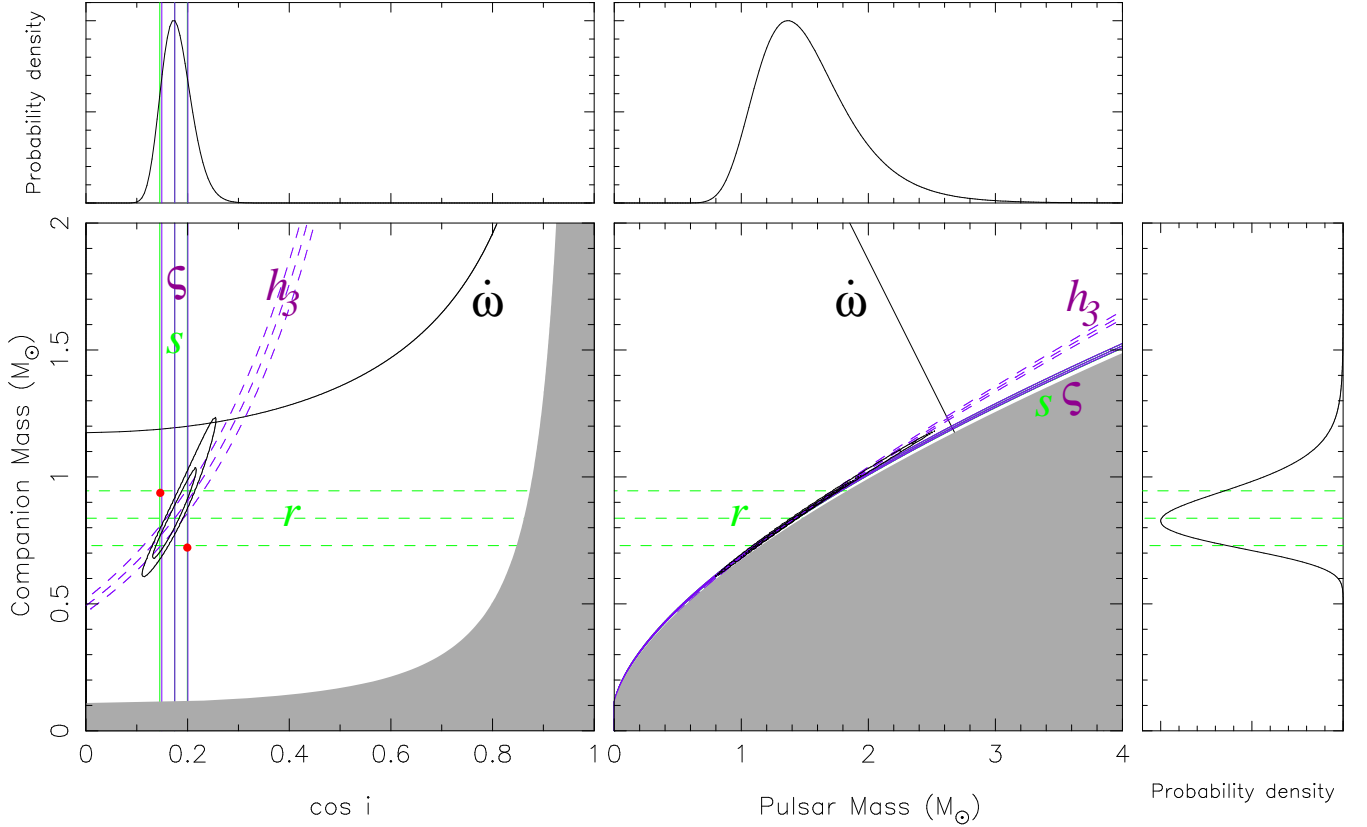}
\caption[]{\small Constraints on the inclination angle and masses in the J1949+3106
  binary system. The black contours include 68.3 and 95.4\% of the total
  probability of a 2-D probability density function (PDF),
  calculated from a $\chi^2$ map of the $h_3$-$h_4$ orthometric space
  that used only the Shapiro delay to constrain the masses.
  The dashed purple lines indicate the constraints from $h_3$ and the
  solid purple lines indicate the constraints from $\varsigma$.
  The solid black line indicates the upper limit derived for $\dot{\omega}$.) 
  The solid and dashed green lines show the
  constraints derived from the $r$-$s$ parameterization of the
  Shapiro delay. The constraint based on $s$ matches that derived from $\varsigma$, as expected. The $r$-$s$
  parameterization is sub-optimal: the points indicated by the
  red dots are 1-$\sigma$ consistent with the values of $r$ and $s$, yet
  they provide very bad fits to the timing data. Nevertheless, both 
  parameterizations provide very similar estimates of $m_{\rm c}$ and
  $\cos i$. {\em left}: $\cos i$-$m_{\rm c}$ plot.
  The gray region is excluded by the condition $m_{\rm p} > 0$.
  {\em Right}: $m_{\rm p}$-$m_{\rm c}$ plot.
  The gray region is excluded by the condition
  $\sin i \leq 1$.   
  {\em Top and right marginal plots}: 1-D PDFs for
  $\cos i$, $m_{\rm p}$ and $m_{\rm c}$, obtained by
  marginalization of the 2-D PDF. From the medians and
  $\pm 1$-$\sigma$ percentiles of these 1-D PDFs, we derive
  $m_p = 1.47^{+0.43}_{-0.31}$~\msun, $m_{\rm c} = 0.85^{+0.14}_{-0.11}$~\msun\ and $i = 79.9^{+1.6}_{-1.9}$ $^\circ$. \label{fig_1949pdfs}}
\end{center}
\end{figure}

\cite{FreireWex10} rewrite the Shapiro delay in terms of two new post-Keplerian parameters: $h_3$, the amplitude of the third orbital harmonic, and $\varsigma$, the ratio of the amplitudes of subsequent harmonics. 
The correlation between the parameters of this ``orthometric'' model is much weaker than the correlation between $r$ and $s$, leading to a better description of the combinations of $m_{\rm c}$ and $\sin i$ allowed by the timing measurements.
We use this variation on the DD binary orbital model for J1949+3106 (Table~\ref{tab_pulsars}). Figure~\ref{fig_1949pdfs} shows the constraints on the pulsar mass, companion mass, and orbital inclination that result from the measurement of $\varsigma$ and $h_3$. It also shows alternative constraints derived using the normal $r$-$s$ parameterization. The figure also shows 68.3\% contours of the 2-D probability density function (PDF) derived from a $\chi^2$ map of the $h_3$-$h_4$ orthometric space, as described in section 5 of Freire \& Wex (2010); this map provides very similar results to a $\chi^2$ map of the $\cos i$-$m_{\rm c}$ space. It is very clear that the 68.3\% contours are better described by the $\varsigma$-$h_3$ parameterization: there are points in the diagram (signaled in red) that are 1-$\sigma$ consistent with $r$ and $s$ that provide bad (i.e., high $\chi^2$) fits to the timing data.

\subsection{Masses of the components of the J1949+3106 binary system}

The lateral panels of Figure~\ref{fig_1949pdfs} show the projection of the 2-D PDF into the $m_{\rm c}$, $\cos i$ and $m_p$ axes. From these 1-D PDFs, we obtain $m_p = 1.47^{+0.43}_{-0.31}$~\msun, $m_{\rm c} = 0.85^{+0.14}_{-0.11}$~\msun\ and $i = 79.9_{+1.6}^{-1.9}$$^\circ$, where uncertainties correspond to $\pm 1$-$\sigma$ confidence levels. 

Intermediate-mass pulsar binaries (IMBP) have C-O or O-Ne-Mg WD companions with $m_{\rm c} > 0.4$~\msun\ \citep{Tauris11b}; for this reason we classify PSR~J1949+3106 as a member of this class. 
\cite{Ferdman10} summarize possible IMBP evolution scenarios and argue that the IMBP J1802$-$2124 has undergone common envelope (CE) evolution, leading to little mass accretion and a history similar to recycled pulsars with neutron star companions. IMBPs with longer orbital periods tend to have larger eccentricities as well \citep{Tauris00}. J1949+3106 is in the orbital period regime where one can argue that CE evolution is likely ($P_{\rm b} < 3$~days).

The observed eccentricities of IMBPs ($e \sim 10^{-4} - 10^{-5}$) are higher on average than those of low-mass pulsar binaries (LMBP), and the orbital eccentricity of PSR~J1949+3106 ($4.3 \times 10^{-5}$) is well within this interval. Furthermore, the shorter accretion episode in IMBPs should not lead to the extreme quenching of the magnetic field seen in LMBPs. This also agrees with observation: IMBPs lie in a region of the $P - \dot{P}$ space distinct from LMBPs \citep{Camilo01}\footnote{They note, however, that this is not consistent with classification based on $m_{\rm c}$: three of the systems have companion masses typical of LMBPs, $0.2 < m_{\rm c} < 0.3$~\msun.}, with higher estimated magnetic fields ($B \propto \sqrt{P~\dot{P}}$). Again, the $P$, $\dot{P}$ and derived $B$ for PSR~J1949+3106 are consistent with those of previously determined IMBPs.


Measuring the mass of PSR~J1949+3106 precisely is important to test our understanding of stellar evolution. For IMBPs, the comparatively short accretion episode has another predictable consequence: the total accreted mass should be (much) smaller than the 0.1-0.5 \msun\ expected for LMBPs \citep{Pfahl02}. Therefore, the masses of pulsars in IMBPs should be very similar to their birth masses. This has recently been confirmed by the mass measurement of the IMBP PSR~J1802$-$2124 ($1.24 \pm 0.11$ \msun) \citep{Ferdman10}. This implies that for J1949+3106 we should also expect a mass close to the Chandrasekhar limit, and the current measurement of $1.47^{+0.43}_{-0.31}~\msun$ is consistent with this expectation. 

A precise measurement of three post-Keplerian parameters ($\varsigma$, $h_3$ and $\dot{\omega}$) will over-determine the mass equations for this system. This implies that this system will eventually provide a test of general relativity.

\subsection{Proper motion}

For pulsars with small period derivatives, the contribution of the Shklovskii effect \citep{Shklovskii70} to $\dot{P}$ can be significant. The Shklovskii effect is due to the change in projected distance between the pulsar and the Solar System barycenter.
Another effect contributing to the observed $\dot{P}$ is due to the difference in acceleration with respect to the Galactic center between the Sun and the pulsar. There is also a small contribution to the observed $\dot{P}$ due to the pulsar being accelerated perpendicularly towards the Galactic plane (\citealt{DT91}, \citealt{NT95}). The contributions of these three effects to the measured $\dot{P}$ values for J1955+2527 and J1949+3106 are given in Table~\ref{tab_pdot}. 
The net value is subtracted from $\dot{P}$ listed in Table~\ref{tab_pulsars} for both pulsars before calculating the estimates for the spin-down luminosity $\dot{E} \propto \dot{P}/P^3$, surface magnetic field $B \propto \sqrt{P \dot{P}}$, and characteristic age $\tau_c = P/2\dot{P}$ in the same table.

One common feature of J1949+3106 and J1955+2527 is that the position angles of both pulsars' proper motions in Galactic coordinates are very closely aligned with the Galactic plane ($\Theta_\mu = 270\degree$ and $262\degree$, respectively, Table~\ref{tab_pulsars}). A major reason for the proper motions being closely aligned with the Galactic plane is Galactic rotation. With the Sun and the pulsar moving around the Galactic center in different directions at $\sim 220$~km~s$^{-1}$, the relative velocity due to this motion is larger than the peculiar motion of the Sun and the motion of the MSP relative to the standard of rest at its position in the Galaxy. Another reason is a selection effect: the PALFA search region is in the Galactic plane, therefore MSPs with a significant component of motion away from the plane are less likely to be found. Given the current locations of the two pulsars ($b = 2.55\degree$ and $-1.58\degree$, respectively), it is unlikely that they will ever move far from the Galactic plane. In this respect they are very similar to PSR~B1855+09 ($z \sim 50$~pc, \citealt{Kaspi94}) and PSR~J1903+0327 ($z < 270$~pc, \citealt{Freire11}), which are also close to the Galactic plane. For the latter system, measurements of the radial velocity of the companion also suggest a relatively small velocity relative to the pulsar's LSR. 

Many ongoing surveys are, for the first time, finding a significant number of MSPs near the Galactic plane (\citealt{Crawford12}, \citealt{HTRUI}, \citealt{HTRUII}, \citealt{HTRUIV}). Measuring the proper motions of these objects will be important to ascertain how many MSPs are tightly confined to the plane of the Galaxy. If there is a statistical excess of such systems then they could represent a separate, low-velocity MSP population.

\section{TOA Error Budget}\label{sec_errorbudget}

In this section we evaluate the relative contributions to the TOA precision that have so far been achieved for J1949+3106 and J1955+2527. First we test whether the residuals are consistent with white noise and then we evaluate the relative contributions to the white noise from three different effects. Given the sharp, unresolved main component in the pulse profile of J1949+3106, it is possible that broadband coherent dedispersion (e.g. with the new PUPPI backend, a GUPPI clone recently installed at Arecibo) might achieve significantly improved timing precision in the near future, providing a much more precise Shapiro delay, much more precise masses and a test of general relativity, as discussed above. 

\subsection{White Noise Tests}

A simple way to test whether the post-fit timing residuals are consistent with white noise is to count zero crossings \citep{Cordes11}. This test is applicable to non-uniformly sampled data and is insensitive to discontinuities in the statistics of the white noise, for example jumps in variance due to using different instruments, as is the case with our residuals. For $N$ samples of white noise, we expect on average $\left<Z_w\right> = \left(N - 1 \right)/2$ zero crossings with a standard deviation of $\sigma_{\rm Z_w} = \sqrt{N - 1}/2$. A comparison between the observed and expected number of zero crossings for the residuals of J1949+3106 and J1955+2527 is shown in Table~\ref{tab_crossings}, including overall and per-observatory results. For 1949+3106, the number of zero crossings is within 1$\sigma$ of the expected white noise value if the data sets from each observatory are treated separately. For J1955+2527, the number of actual and expected zero crossings is within 1$\sigma$ for GBT TOAs and within 2$\sigma$ for all other observatories. 

Another white noise test involves fitting for a second frequency or period derivative. We expect that to be consistent with zero for white-noise-like residuals, and this is the result we obtain for J1949+3106. For J1955+2527, we detect a second frequency derivative of $-2.2(7)$~s$^{-3}$, confirming the indication from the zero crossing tests that this pulsar's residuals may exhibit some red noise characteristics.

\subsection{Template Fitting Error}

Times of arrival are obtained by folding small portions of an observation (typically a few minutes) with an ephemeris that describes the timing solution of a pulsar, and then convolving the resulting folded profile with a pulse shape template. The pulse shape template used for extracting TOAs at a given observing frequency is the average of the folded profiles of many observations of the same pulsar at the same frequency. This is the method we use for obtaining TOAs for J1949+3106 and J1955+2527. 

Since TOA extraction is based on matched filtering, we can view the TOA and its uncertainty as a measurement of the pulse shape, which is affected both by the quality of the template and the amount of noise in the data. If the pulse shape does not change between observations, it can be described as an invariant template added to noise. Assuming that the noise is white, \cite{Cordes10} derive the minimum TOA error, $\sigma_{t_{\rm S/N}}$, from system parameters and an effective pulse width $W_{\rm eff}$. 


For a noiseless Gaussian pulse, $W_{\rm eff} = 0.6~{\rm FWHM}$. Using Equation~A2 from \cite{Cordes10} and smoothed versions of the TOA extraction templates for Arecibo observations of both pulsars with 1024 bins in the folded profile, we obtain $W_{\rm eff} = 90~\mu$s for J1949+3106 and $W_{\rm eff} = 109~\mu$s for J1955+2527. For comparison, ${\rm FWHM} \sim 560~\mu$s for J1955+2527, with unresolved structure on the leading edge of the pulse deviating from a Gaussian. J1949+3106 has a two-component main pulse with the FWHM of the brighter (and narrower) component of the main pulse $\approx 180~\mu$s. In both cases, $W_{\rm eff} < 0.6~{\rm FWHM}$ of the main pulse, because the TOA precision depends on the width of the narrowest substructure in the pulse that the template can match. In the case of J1949+3106, where the two components of the main pulse are well resolved, $W_{\rm eff} \approx 0.6~{\rm FWHM}$ of the narrower component. For J1955+2527, $W_{\rm eff}$ is about a third of $0.6~{\rm FWHM}$. Even though the pulse profile of this pulsar is Gaussian-like, it is not completely featureless. The calculated value of $W_{\rm eff}$ is likely affected by the presence of an unresolved bump on the leading edge of the main pulse and/or a slight bump at the very top of the pulse. 

For Arecibo L-band timing observations, $T_{\rm sys} = 30$~K and $G = 10$~K/Jy, giving $S_{\rm sys} = 3$~Jy. The bandwidth per WAPP was 50~MHz, and each TOA was obtained by folding 500~s of data. The peak pulse flux $S_{\rm peak} \sim 3.2$~mJy for J1949+3106 and $\sim 1.8$~mJy for J1955+2527. Using Equation~A1 from \cite{Cordes10}, we obtain $\sigma_{t_{\rm S/N}} = 4.6~\mu$s for J1949+3106 and $\sigma_{t_{\rm S/N}} = 5.3~\mu$s for J1955+2527.



\subsection{Pulse Jitter}


The folded pulse profile used in TOA extraction is obtained from averaging many pulses. For pulsars in general, each individual pulse is narrower than the average pulse profile, and the average pulse profile extends over the phase window where individual pulses are observed. Individual pulse phases may vary by an amount on the order of a pulse width from one pulse to the next. The folded pulse profile from which a TOA is extracted depends on both the shapes of individual pulses and the distribution of their phases. 

The intensity modulation index $m_{I} = \sigma_{\rm S} / \left<S\right>$ (the ratio of intensity rms and mean as a function of pulse phase) is used to characterize the amplitude modulation and the phase jitter and is typically of order unity (e. g. \citealt{Helfand77}, \citealt{Bartel80}, \citealt{Weisberg86}). \cite{Cordes10} show that the TOA error due to pulse phase jitter, $\sigma_{t_{\rm J}}$, can be expressed in terms of $m_{I}$, the intrinsic pulse width $W_{\rm int}$, and a factor $f_{\rm J}$.
Previous studies of profile stability (\citealt{Helfand75}, \citealt{rr95}) show results that are consistent with  $f_{\rm J} \sim 1/3 - 1/2$ for most pulsars. Using Equation~A6 from \cite{Cordes10}, we calculate an upper limit on $\sigma_{t_{\rm J}}$ by assuming $m_{I} = 1$, $f_{\rm J} = 1/2$, and $W_{\rm int}$ equal to the FWHM of the 1.4~GHz template used to extract the Arecibo TOAs. 
With these parameters, we get $\sigma_{t_{\rm J}} = 1.2~\mu$s for J1949+3106 and $\sigma_{t_{\rm J}} = 0.52~\mu$s for J1955+2527. 

The average signal-to-noise ratio of an individual pulse is ${\rm SNR_1 = SNR_{\rm prof}}/\sqrt{N_{\rm pulses}}$, where $\rm SNR_{\rm prof}$ is the signal-to-noise ratio of the folded TOA profile and $N_{\rm pulses}$ is the number of pulses averaged to produce the TOA profile. Generally, if $\rm SNR_1 > 1$, the TOA uncertainty is dominated by pulse jitter as opposed to radiometer noise. Using our Arecibo TOA profiles, we obtain $\rm SNR_1 \sim 0.04$ for J1949+3106 and $\rm SNR_1 \sim 0.01$ for J1955+2527, consistent with the above result that $\sigma_{t_{\rm J}} < \sigma_{t_{\rm S/N}}$ and therefore the TOAs are radiometer noise dominated for both pulsars. 

\subsection{Diffractive Scintillation}

The characteristics of scintles in any given portion of data that is processed to produce a TOA affect the folded pulse profile and introduce an error in the TOA \citep{Cordes90}. We calculate this error for J1949+3106 and J1955+2527 from an estimate of the scattering broadening time $\tau_s$ at 1.4~GHz. 

An upper bound on the scattering time is simply $W_{\rm eff}$; therefore we have $\tau_s < 90~\mu$s for J1949+3106 and $\tau_s < 109~\mu$s for J1955+2527. For comparison, we use the NE2001 model of the ionized gas distribution in the galaxy \citep{NE2001} for the coordinates and dispersion measures of the two pulsars to obtain $\tau_s$ at 1~GHz and scale it to 1.4~GHz assuming a Kolmogorov scattering spectrum with an index of $-4.4$. For an observing frequency of 1.4~GHz, we obtain $\tau_s = 0.84~\mu$s for J1949+3106 and 1.6~$\mu$s for J1955+2527, consistent with our upper limits. As a further comparison, we use the empirical relation derived by \cite{Bhat04}, who assemble a set of pulsars with DM and $\tau_{\rm s}$ measurements and fit a parabola with log(DM) as the independent variable.
This relation gives $\tau_s = 8~\mu$s for J1949+3106 and 127~$\mu$s for J1955+2527, though it is worth noting that there is scatter of up to two orders of magnitude in $\tau_{\rm s}$ about the fit. 

The scintillation bandwidth $\Delta f_{\rm DISS}$ and time scale $\Delta t_{\rm DISS}$ could not be constrained from the data since the two pulsars are not bright enough to observe scintles. We find $\Delta f_{\rm DISS}$ following \cite{Lambert99}. We calculate $\Delta t_{\rm DISS}$ using Equations 11 and 12 from \cite{Cordes98} and the distance $D$ and transverse velocity $V_T$ for each pulsar (Table~\ref{tab_pulsars}). 
Following \cite{Cordes10}, we calculate the number of scintles in time ($N_{\rm t}$) and frequency ($N_{\rm f}$) for the portion of data used to produce each TOA.
We obtain $N_{\rm t} = 1.9$ and $N_{\rm f} = 51$ for J1949+3106, and $N_{\rm t} = 2.2$ and $N_{\rm f} = 101$ for J1955+2527. Finally, we calculate the rms error in the scattering broadening function due to scintillation, $\sigma_{t_{\delta DISS}}$, from Equation~23 in \cite{Cordes10}. 
Using the upper bounds on $\tau_{\rm s}$ from the FWHM pulse width in the above calculations, we obtain $\sigma_{t_{\delta DISS}} < 0.09~\mu$s for J1949+3106 and $\sigma_{t_{\delta DISS}} < 0.07~\mu$s for J1955+2527.

\subsection{Dispersion Measure Variations}\label{section_DM}


The dispersion measures of J1949+3106 and J1955+2527 and their uncertainties (Table~\ref{tab_pulsars}) were determined by TEMPO2 fits to TOAs extracted from several frequency subbands per observation. While we have not observed DM variations in J1949+3106 and J1955+2527, we can place an upper limit on the TOA error due to unmodeled DM variations (whose effect could be partly or fully absorbed by the other parameters in the timing model) based on the DM uncertainties for both pulsars from the timing solutions in Table~\ref{tab_pulsars}. The maximum contribution to the TOA uncertainty from unmodeled DM variations is
\be
\sigma_{t_{\rm DM, max}} = \frac{8.3{\rm \mu s}~\delta {\rm DM}~\Delta\nu}{f_{\rm center}^3},
\ee
where $\Delta\nu$ is the bandwidth in MHz, $f_{\rm center}$ is the center observing frequency in GHz, and $\delta \rm DM$ is twice the DM uncertainty. At the lowest frequency Arecibo observations used, 1420~MHz, the bandwidth of 50~MHz implies DM uncertainties of 0.0006 and 0.003~pc~cm$^{-3}$ for J1949+3106 and J1955+2527, respectively. These values give $\sigma_{t_{\rm DM, max}} = 0.2~\mu$s for J1949+3106 and 0.9~$\mu$s for J1955+2527. 

\cite{Backer93} fit observed gradients of pulsar DMs and find that the annual DM gradient is proportional to $\sqrt{DM}$. Based on their Figure~4a, we estimate an annual DM gradient of $\sim 0.002$~pc~cm$^{-3}$ for J1949+3106 and $\sim 0.003$~pc~cm$^{-3}$ for J1955+2527. For the $\sim 4$ years that we have been timing both pulsars, this empirical relation predicts a DM gradient contribution to the rms residual of 1.2 and 1.8~$\mu$s, respectively. Both predictions are within an order of magnitude of our estimates based on the DM uncertainties reported by TEMPO2 above.

\section{Summary and Conclusions}

To the current tally of almost 2000 known pulsars we have added two MSPs found by the PALFA survey, J1949+3106 and J1955+2527, and presented their timing solutions. While J1955+2527 is isolated, in the J1949+3106 binary system we have been able to confidently measure the Shapiro delay and estimate the pulsar and companion masses. The pulsar's current median mass is 1.47~\msun, and the companion's median mass is 0.85~\msun. The uncertainties of these mass estimates are still $0.3 - 0.4~\msun$ and $0.1 - 0.2~\msun$, respectively, but they will improve with continued timing. We are also on the verge of being able to detect the relativistic periastron advance in this system. 

We have outlined the steps towards breaking down the various contributions to the overall rms timing residual and applied them to the TOA sets of the two new pulsars. This type of characterization is important with a view to figuring out what effects and in what cases are mostly responsible for the observed residuals, and what can we do to mitigate them. This has further implications for how we can take maximum advantage of the properties of new discoveries in projects like the International Pulsar Timing Array (IPTA), which need extremely precise timing measurements on pulsars that are very stable natural clocks. The North American Nanohertz Observatory for Gravitational Waves (NANOGrav\footnote{\texttt{http://nanograv.org}}), an IPTA member, uses the Arecibo telescope for timing pulsars suitable for the IPTA and is especially interested in new MSP discoveries in the Arecibo sky. In the case of J1949+3106 and J1955+2527, the rms timing residuals are too large for including these pulsars in the IPTA sample. Table~\ref{tab_toaerror} lists overall rms timing residuals for both pulsars, residuals by observatory, the expected contribution to the residual from radiometer noise, and upper limits on the contributions from pulse jitter, diffractive scintillation, and unmodeled DM variations. We find that radiometer noise puts a hard limit on TOA precision in both cases. The overall timing residual of J1949+3106 is consistent with that limit, while the residual of J1955+2527 is more than twice as large. The TOA residuals of J1949+3106 are consistent with white noise, while J1955+2527 exhibits some modest departures from white noise as evidenced in a zero-crossing test and in a fit for a second frequency derivative.

\section{Acknowledgments}

The Arecibo Observatory is operated by SRI International under a cooperative agreement with the National Science Foundation (AST-1100968), and in alliance with Ana G. Ana G. M\'{e}ndez-Universidad Metropolitana, and the Universities Space Research Association. The Nan\c{c}ay radio telescope is part of the Paris Observatory, associated with the Centre National de la Recherche Scientifique (CNRS), and partially supported by the Region Centre in France. This work was supported by NSF grant AST-0807151 to Cornell University. PALFA research at UBC is funded by NSERC, the Canada Foundation for Innovation, and CANARIE. PF gratefully acknowledges the financial support by the European Research Council for the ERC Starting Grant BEACON under contract no. 279702. JWTH is a Veni Fellow of the Netherlands Foundation for Scientific Research (NWO). BK gratefully acknowledges the support of the Max Planck Society. PL was partly funded for this research by an NSERC PGS scholarship, and an IMPRS fellowship.

\begin{deluxetable}{llcc}
\tablecolumns{4}
\tablewidth{0pc}
\tablecaption{Contributions to TOA uncertainty (all values in $\mu$s).\label{tab_toaerror}}
\tablehead{
\colhead{Parameter} & \colhead{Observatory} & \colhead{PSR J1949+3106} & \colhead{PSR J1955+2527}}
\startdata
Timing residual & Arecibo & \phn\phn3.7 (1119) &  \phn\phn7.9 (199) \\
(Number of TOAs) & GBT & 12.8 (56)\phn & 58.2 (97) \\
 & Jodrell & 17.1 (85)\phn & \phn36.2 (101)\\
 & Nan\c{c}ay & 4.3 (28) & 11.8 (16) \\
\\
EFAC\tablenotemark{a} & Arecibo & 0.84 & 0.96 \\
& GBT & 1.09 & 1.61 \\
& Jodrell & 1.20 & 1.22 \\
& Nan\c{c}ay & 1.24 & 5.70\\
\\
Overall rms timing residual & & 3.96 & 11.4 \\
\\
Radiometer noise, $\Delta t_{\rm S/N}$ & Arecibo & 4.6 & 5.3 \\
Pulse jitter, $\Delta t_{\rm J}$ & Arecibo & $<1.2$ & $<0.52$ \\
Diffractive scintillation, $\Delta t_{\delta DISS}$ & Arecibo & $<0.09$ &  $<0.07$\\
Unmodeled DM variations, $\Delta t_{\rm DM}$  & & & \\
\phn\phn From DM uncertainty\tablenotemark{b} & Arecibo & $< 0.2$ & $< 0.9$ \\
\phn\phn From DM gradient fit\tablenotemark{c} & Arecibo & $< 1.2$ & $< 1.8$ \\

\enddata
\tablenotetext{a}{Weighting factor for TOA uncertainties chosen such that the ratio of reduced $\chi^2$ in the TEMPO2 fit to the number of degrees of freedom is $\sim 1$.}
\tablenotetext{b}{This paper, Table~\ref{tab_pulsars}}
\tablenotetext{c}{\cite{Backer93}, Figure 4a}

\end{deluxetable}

\begin{deluxetable}{lcc}
\small
\tablecolumns{3}
\tablewidth{0pc}
\tablecaption{Pulsar parameters.\label{tab_pulsars}}
\tablehead{
\colhead{Parameter} & \colhead{PSR J1949+3106} & \colhead{PSR J1955+2527}}
\startdata
Right ascension (J2000) & 19$^h$~49$^m$~29.63851$^s$(1) &  19$^h$~55$^m$~59.39523$^s$(7)  \\
Declination (J2000) & 31\degree~06\amin~03.8289\asec(3) &  25\degree~27\amin~03.443\asec(2) \\
Proper motion in RA, $\mu_\alpha$ (mas/yr) & $-2.94(6)$ & $-1.9(6)$ \\
Proper motion in DEC, $\mu_\delta$ (mas/yr) & $-5.17(8)$ & $-2.4(8)$ \\
Spin frequency, $f$ (s$^{-1}$) & 76.114023821963(3) & 205.22225531037(7) \\
Frequency derivative, $\dot{f}$ (s$^{-2}$) & $-5.4407(3) \times 10^{-16}$ & $-3.84(4) \times 10^{-16}$\\
Second frequency derivative, $\ddot{f}$ (s$^{-3}$) & - & $-2.2(7) \times 10^{-25}$\\
Epoch of timing solution (MJD) & 54500.000176077 & 54800.0 \\
Dispersion measure, DM (pc~cm${}^{-3}$) & 164.1264(5) & 209.971(3) \\

\\
Orbital period, $P_{b}$ (days) & 1.949535(2) & - \\
Time of periastron passage, $T_{0}$ (MJD) & 54390.270(1) & - \\
Projected semi-major axis, $x$ (lt-s)\tablenotemark{a} & 7.288650(1) & - \\
Longitude of periastron, $\omega$ & 207.5(2)\degree & - \\
Eccentricity & 0.00004306(7) & - \\
Orthometric harmonic amplitude, $h_3$ & $2.4(1) \times 10^{-6}$ & - \\
Orthometric harmonic ratio, $\varsigma$ & 0.84(2) & - \\
\\

Period, $P$~(s) & 0.0131381833437039(5) & 0.004872765862979(2)  \\
Period derivative, $\dot{P}$~(s~s$^{-1}$) & 9.3913(5) $\times 10^{-20}$ & 9.12684(8) $\times 10^{-21}$  \\
Mass function (\msun) & 0.10939(5) & - \\
Pulsar mass (\msun) & $1.47_{-0.31}^{+0.43}$ & - \\
Companion mass (\msun) & $0.85_{-0.11}^{+0.14}$ & - \\
Orbital inclination, $i$ (deg) & $79.9^{+1.6}_{-1.9}$  & - \\
Rate of periastron advance, $\dot{\omega}$ (deg/yr) & $< 0.02$ & - \\
\\
Galactic longitude, $l$ (deg) & 66.86 & 62.74 \\
Galactic latitude, $b$ (deg) & $+$2.55 & $-$1.58 \\
Total proper motion, $\mu$ (mas/yr) &  5.9(1) & 3.1(8) \\
Position angle of proper motion, $\Theta_{\mu}$ (J2000) & 211(1) & 218(13) \\
Position angle of proper motion, $\Theta_{\mu}$ (Galactic) & 270(1) & 276(13) \\
Surface magnetic field, $B$ (G)\tablenotemark{b} &  $1.12\times 10^{9}$ & $2.14\times 10^{8}$ \\
Spin-down luminosity, $\dot{E}$ (erg~s$^{-1}$)\tablenotemark{c} & $1.63 \times 10^{33}$ & $3.21 \times 10^{33}$ \\
Characteristic age, $\tau_{c}$ (Gyr)\tablenotemark{d} & 2.2 & 8.3 \\
Distance, $D$ (kpc)\tablenotemark{e} & 6.5 & 7.5 \\
Transverse velocity, $V_{T}$ (km~s$^{-1}$) & $\sim 180$ & $\sim 107$  \\
Distance from Galactic plane, $|z|$, (kpc)\tablenotemark{f} & 0.29 & 0.21 \\
1400~MHz flux density, $S_{1400}$ (mJy) & $0.23\pm0.05$ & $0.28\pm0.06$\\
1400~MHz radio luminosity, $L_{1400}$ (mJy~kpc$^2$)\tablenotemark{g} & $\sim$~9.7 & $\sim$~15.8 \\
\\
Number of points in timing fit & 1288 & 413 \\
Weighted rms post-fit residual ($\mu$s) & 3.96 & 11.4 \\
Timing span (d) & 1715 & 1773 \\
\enddata
\tablenotetext{a}{$x = a\ sin\ i/c$, where $a$ is the semi-major axis and $i$ is the inclination angle.}
\tablenotetext{b}{$B = 3.2 \times 10^{19}~(P\dot{P})^{1/2}$ }
\tablenotetext{c}{$\dot{E} = 4 \pi^2 I \dot{P} / P^3$ and assuming a 1.4~\msun\ neutron star with a 10~km radius and moment of inertia $I = 10^{45}$~g~cm$^{-3}$.}
\tablenotetext{d}{$\tau_{c} = P / 2\dot{P}$}
\tablenotetext{e}{Estimates based on DM, sky position, and the NE2001 model of ionized gas in the Galaxy \citep{NE2001}.}
\tablenotetext{f}{$|z| = D~\rm{sin}~|b|$}
\tablenotetext{g}{$L_{1400} = S_{1400}~D^2$}
\end{deluxetable}

\begin{deluxetable}{lrr}
\tablecolumns{3}
\tablewidth{0pc}
\tablecaption{Contributions to measured $\dot{P}$ values.\label{tab_pdot}}
\tablehead{
\colhead{Effect} & \colhead{PSR J1949+3106} & \colhead{PSR J1955+2527}}
\startdata
Shklovskii effect (s~s$^{-1}$) & $7(2) \times 10^{-21}$ & $2.6(5) \times 10^{-21}$\\
Acceleration $\parallel$ to Galactic plane (s~s$^{-1}$) & $-6(1) \times 10^{-21}$ &  $-2.9(4) \times 10^{-21}$ \\
Acceleration $\perp$ to Galactic plane (s~s$^{-1}$) & $-7.8(7) \times 10^{-23}$ & $-1.5(2) \times 10^{-23}$ \\
Total $\dot{P}$ contribution (s~s$^{-1}$) & $0.274 \times 10^{-21}$ & $ -0.278 \times 10^{-21}$ \\
\enddata
\end{deluxetable}

\begin{deluxetable}{lrrrr}
\tablecolumns{5}
\tablewidth{0pc}
\tablecaption{Number of TOAs, actual number of residual zero crossings, expected number of zero crossings for white-noise-like residuals, and expected standard deviation of the latter per observatory for J1949+3106 and J1955+2527.\label{tab_crossings}}
\tablehead{
\colhead{Observatory} & \colhead{TOAs} & \colhead{$Z$} & \colhead{$\left<Z_w\right>$} & \colhead{$\sigma_{\rm Z_w}$}}
\startdata
\emph{J1949+3106} & & & & \\
\phn \phn Arecibo & 1119 & 551 & 559 & 17 \\
\phn \phn GBT & 56 & 31 & 28 & 4 \\
\phn \phn Jodrell Bank & 85 & 42 & 42 & 5 \\
\phn \phn Nan\c{c}ay & 28 & 16 & 14 & 3 \\
\phn \phn All & 1288 & 613 & 612 & 17 \\
\\
\emph{J1955+2527} & & & & \\
\phn \phn Arecibo & 199 & 109 & 99 & 7 \\
\phn \phn GBT & 97 & 49 & 48 & 5 \\
\phn \phn Jodrell Bank & 101 & 43 & 50 & 5 \\
\phn \phn Nan\c{c}ay & 16 & 4 & 8 & 2 \\
\phn \phn All & 413 & 206 & 206 & 10 \\
\enddata
\end{deluxetable}

\cleardoublepage

\end{document}